\documentclass[12pt]{article}
\usepackage{amsfonts}
\usepackage{amsmath}
\usepackage{amssymb}
\usepackage{amstext}
\usepackage{graphicx,epsfig}
\usepackage{hyperref}
\usepackage[nooneline]{caption2}
\makeatletter
\def\ExtendSymbol#1#2#3#4#5{\ext@arrow 0099{\arrowfill@#1#2#3}{#4}{#5}}
\def\RightExtendSymbol#1#2#3#4#5{\ext@arrow 0359{\arrowfill@#1#2#3}{#4}{#5}}
\def\LeftExtendSymbol#1#2#3#4#5{\ext@arrow 6095{\arrowfill@#1#2#3}{#4}{#5}}

\makeatother

\begin{document}
\baselineskip 20pt

\title{Proposal for a new scheme for producing a two-photon, high dimensional hyperentangled state}
\author{Hong-Bo Xu$^a$, Kun Du$^a$ and
Cong-Feng Qiao$^{a,b}$\footnote{Corresponding author. Email: qiaocf@gucas.ac.cn}\\[0.5cm]
$^{a}$Department of Physics, Graduate University of
Chinese Academy of Sciences,\\ Beijing 100049, China\\[0.2cm]
$^{b}$Theoretical Physics Center for Science Facilities (TPCSF),\\
CAS, Beijing 100049, China}

\date{}
\maketitle

\begin{abstract}

We propose an experimentally feasible scheme for generating a two
$2\times4\times4$ dimensional photons hyperentangled state,
entangled in polarization, frequency and spatial mode. This scheme
is mainly based on a parametric down-conversion source and
cross-Kerr nonlinearities, which avoids the complicated uncertain
post-selection. Our method can be easily expanded to the production
of hyperentangled states with more photons in multidimensions. Hence
the expectation for vast quantities of information in quantum
information processing will possibly come true. Finally, we put
forward a realizable quantum key distribution (QKD) protocol based
on the high dimensional hyperentangled state.
\end{abstract}

\section{Introduction}

Entanglement is viewed as a kind of raw resource of quantum
information science, such as measurement-based quantum computing
\cite{one-way quantum computer}, quantum teleportation
\cite{Teleporting}, quantum dense coding \cite{C. Wang et al},
entanglement purification \cite{Purification of Noisy Entanglement}
and quantum cryptography \cite{D. Bruss and C. Macchiavello}. Thus
far there have been plentiful protocols in quantum information
processing by applying two-dimensional quantum systems-qubits. Due
to the demand for more and more information content, some research
has focused on extending qubits to multidimensional entangled
states-qudits. In recent years, three-dimensional quantum states
(qutrits) \cite{Three-Dimensional Entanglement,Triggered qutrits}
and higher dimensional quantum states \cite{qudits using twin
photons,engineered two-photon,Pixel Entanglement} have been experimentally realized. The qudits have shown better
characteristics and advantages than qubits. For instance,
multidimensional entanglement has shown stronger quantum
nonlocality \cite{Violations of Local Realism} and noise immunity
\cite{Arbitrarily High-Dimensional Systems}. Moreover, qudits can
observably improve the security of quantum key distribution
\cite{3-State Systems,entangled qutrits,multilevel encoding,d-Level
Systems}.

Notwithstanding at present the relevant methods to prepare a state
in arbitrary d-dimensional Hilbert space entangled in one degree of
freedom (DOF) have been proposed, the experimental realization of
multipartite entanglement is still a significant challenge. Being
analogous to the case of qubits, hyperentanglement
\cite{hyperentanglement} provides an effective and practical
capacity-increased way to manipulate more qudits in arbitrary
desirable Hilbert dimensions. Furthermore, hyperentanglement is much
less affected by decoherence and plays an important role in the
realization of even more challenging quantum information processing
in comparison with the normal entangled states.

In this paper, we propose a scheme for generating a two
$2\times4\times4$ dimensional photon hyperentangled state,
entangled in polarization, frequency and spatial mode DOFs£¬
respectively. After that we discuss a feasible quantum key
distribution protocol based on this hyperentangled state. Our
protocol can observably increase the efficiency of key distribution
and the flux of information.

\section{Preparation scheme for multidimensional hyperentangled state}

In this section, a preparation scheme of the two $2\times4\times4$
dimensional photon hyperentangled state, entangled in polarization,
frequency and spatial mode, is introduced. By means of the scheme,
the two photons are simultaneously entangled in polarization,
frequency and spatial modes, encoded in two-dimensional, four-dimensional
and four-dimensional Hilbert spaces respectively. The hyperentangled
state takes the following form up to a normalization constant:
\begin{eqnarray}
|\psi\rangle=(HV+VH)(\omega_{11}\omega_{12}+
\omega_{12}\omega_{11}+\omega_{21}\omega_{22}+
\omega_{22}\omega_{21})(a_{11}b_{11}+a_{12}b_{12}+
a_{21}b_{21}+a_{22}b_{22}) ,\label{shi1}
\end{eqnarray}
where $H$ and $V$ denote horizontal and vertical polarization,
$\omega_{11}$, $\omega_{12}$, $\omega_{21}$, $\omega_{22}$ signify
different frequencies and $a_{11}$, $a_{12}$, $a_{21}$, $a_{22}$ and
$b_{11}$, $b_{12}$, $b_{21}$, $b_{22}$ label different spatial
modes.

The first step is to produce primary light source which has the form
as $|\omega_{1}\rangle+|\omega_{2}\rangle$ ($\omega_{2}=4\omega_{1}$) in frequency. The spatial mode of
the Initial laser pulse is turned into a superposition of two new
spatial modes by a 50:50 beam splitter (BS). Put two frequency
multipliers (FM) \cite{frequency multiplier} on the above path, and
then the two paths are coupled with another BS. Right now the
superposition of the two spatial modes has turned into a
superposition of two new frequencies.

\begin{figure}[htp]
\centering
\includegraphics[width=0.8\textwidth]{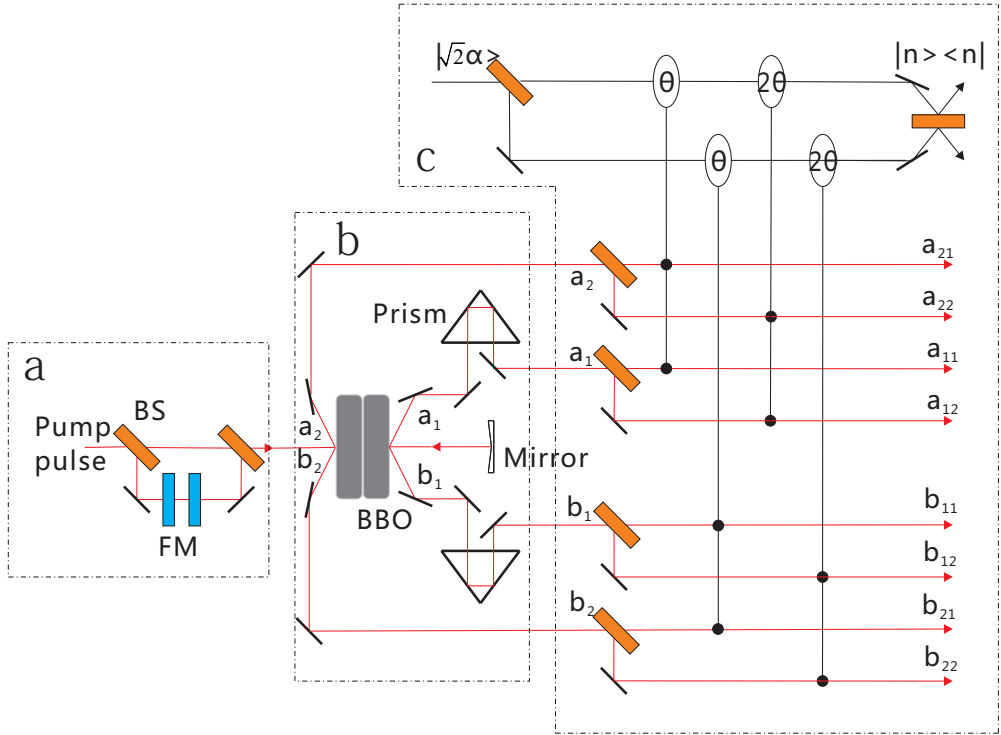}
\caption{\small Preparation scheme of the two $2\times4\times4$
dimensional photons hyperentangled state via type-II barium borate
(BBO) crystal and cross-Kerr nonlinearity.} \label{fig1}
\vspace{-0mm}
\end{figure}

Then as shown in Figure \ref{fig1} through two adjacent type-II barium borate
(BBO) crystals \cite{BBO}, for the first (second) crystal, the ratio of the frequency
of the signal photon to the frequency of the idler
photon is 1:2 (2:1) \cite{hyperentanglement}, thus the photon pairs can be encoded in
two-dimensional polarization entanglement and four-dimensional frequency
entanglement
($|\omega_{1}\rangle$=$|\omega_{11}\rangle$+$|\omega_{12}\rangle$,
$|\omega_{2}\rangle$=$|\omega_{21}\rangle$+ $|\omega_{22}\rangle$,
$\omega_{11}$:$\omega_{12}$:$\omega_{21}$:$\omega_{22}$=1:2:4:8) by
spontaneous parametric down-conversion (SPDC) \cite{SPDC} and
emitted into the spatial modes $a_{1}$ and $b_{1}$. In case the pump
pulse passes through the crystal with no photon pairs emitted, it
will be reflected by the mirror and transit the crystal a second
time, which may produce corresponding photon pairs in spatial modes
$a_{2}$ and $b_{2}$ \cite{Purification using Spatial Entanglement}.
As a result, we encode the two-photon state in two-dimensional spatial
mode entanglement as well.  After eliminating the undesired timing
information \cite{Bell-state Synthesizer},
the two-photon hyperentangled states in the following form are
readily obtained:
\begin{eqnarray}
|\psi\rangle=(HV+VH)(\omega_{11}\omega_{12}+
\omega_{12}\omega_{11}+\omega_{21}\omega_{22}+
\omega_{22}\omega_{21})(a_{1}b_{1}+a_{2}b_{2}) .\label{shi2}
\end{eqnarray}

Next, let photons in every path enter a BS, so that every spatial mode
is divided into two new spatial modes with equal probability and
fixed phase relation. Since one can simply adjust the relative
phase, the two-photon state in the new eight spatial modes can be
expressed as
$(a_{11}+a_{12})(b_{11}+b_{12})+(a_{21}+a_{22})(b_{21}+b_{22})$.

Afterwards, these eight paths are led to a cross-kerr nonlinear
medium \cite{cross-kerr,quantum communication}, which brings forth
an adjustable phase shift to the coherent state through cross-phase
modulation (XPM). First using a BS to divide the coherent state into
two beams $|\alpha\rangle$ $|\alpha\rangle$, adjust the phase shifts
of the upper $|\alpha\rangle$ into $\theta$, $\theta$, $2\theta$ and
$2\theta$ for modes $a_{11}$, $a_{21}$, $a_{12}$ and $a_{22}$
respectively. Analogously, modes $b_{11}$, $b_{21}$, $b_{12}$ and
$b_{22}$ induce the phase shifts of the under $|\alpha\rangle$ into
$\theta$, $\theta$, $2\theta$ and $2\theta$ respectively. Then the
two coherent states are compared with a BS. After projecting the
$|n\rangle$$\langle n|$ onto the upper beam \cite{homodyne,double
XPM}, if $n=0$, the state in spatial mode transforms into
$a_{11}b_{11}+a_{12}b_{12}+a_{21}b_{21}+a_{22}b_{22}$, i.e.
four-dimensional spatial mode entanglement. Finally, the state (2) has
turned into the state (1),  the two $2\times4\times4$ dimensional
photons hyperentangled state is created.

\begin{figure}[htp]
\centering
\includegraphics[width=0.5\textwidth]{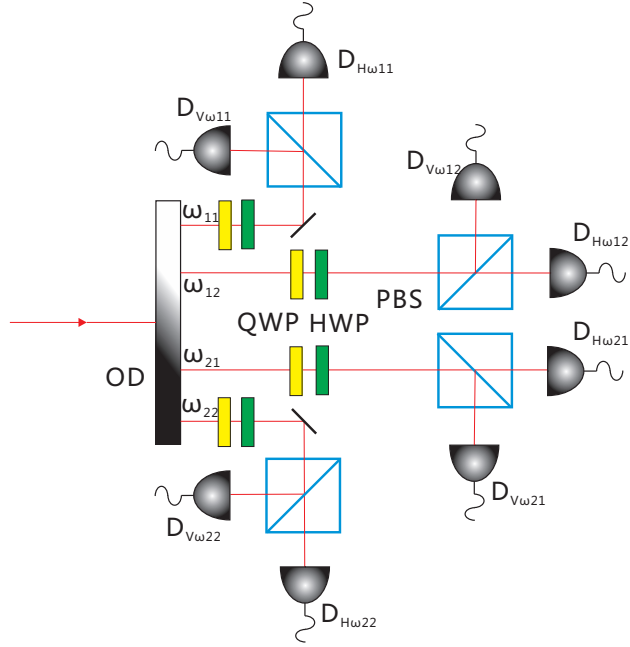}
\caption{\small The setup for measuring the high dimensional
hyperentangled state.} \label{fig2} \vspace{-0mm}
\end{figure}

A setup for measuring the hyperentanglement in polarization,
frequency and spatial mode simultaneously and independently is
schematically shown in Figure \ref{fig2}. Place eight of these
setups in all eight output modes, then the qudits in spatial mode can be
determined. Then every mode is split into four paths according to
frequency by an optical demultiplexer (OD) \cite{WDM}, hence the
qudits in frequency  are obtained. Thereafter the conventional
polarization analysis \cite{polarization analysis} is applied in all
the four paths in order to read out the qubits in polarization.

As the theory goes, our scheme can be simply expanded to the
production of a hyperentangled state with an arbitrary number of
photons and arbitrary dimensional entanglement, which will immensely
increase the quantity of information.

\section{QKD protocol for multidimensional hyperentangled state}

Now we shall present a QKD protocol using the high dimensional
hyperentangled state generated above. In our method, we utilize the
four-dimensional hyperentanglement swapping \cite{hyperentanglement
swapping} in frequency and spatial mode DOFs of the two photons lying
in the state of following form:
\begin{eqnarray}
|\psi\rangle=(|\omega_{11}\rangle|\omega_{12}
\rangle + |\omega_{12}\rangle|\omega_{11}\rangle+
|\omega_{21}\rangle|\omega_{22}\rangle+
|\omega_{22}\rangle|\omega_{21}\rangle)\nonumber\\
(|a_{11}\rangle|b_{11}\rangle+
|a_{12}\rangle|b_{12}\rangle+
|a_{21}\rangle|b_{21}\rangle+|a_{22}\rangle|b_{22}\rangle).
\label{shi3}
\end{eqnarray}

\begin{figure}[htp]
\centering
\includegraphics[width=0.6\textwidth]{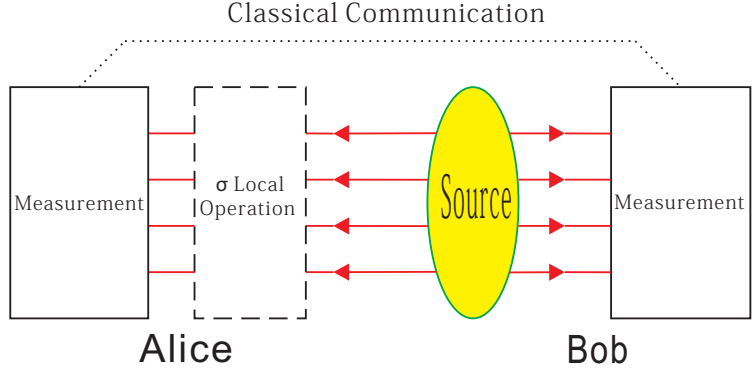}
\caption{\small Scheme showing the principle of QKD protocol using
high dimensional hyperentangled state.} \label{fig3} \vspace{-0mm}
\end{figure}

As shown in Figure \ref{fig3}, send the two photons to Alice and Bob
respectively, named photon A and photon B. Then the state (3) can be
expressed as:
\begin{eqnarray}
|\psi\rangle=\frac{1}{8}(|\psi_{11}\rangle_{A}|
\psi_{11}\rangle_{B}+|\psi_{12}\rangle_{A}|
\psi_{12}\rangle_{B}+|\psi_{13}\rangle_{A}|
\psi_{13}\rangle_{B}+|\psi_{14}\rangle_{A}|
\psi_{14}\rangle_{B}\nonumber\\
                   +|\psi_{21}\rangle_{A}|
                   \psi_{21}\rangle_{B}+|
                   \psi_{22}\rangle_{A}|
                   \psi_{22}\rangle_{B}+
                   |\psi_{23}\rangle_{A}|
                   \psi_{23}\rangle_{B}+|
                   \psi_{24}\rangle_{A}|
                   \psi_{24}\rangle_{B}\nonumber\\
                   +|\psi_{31}\rangle_{A}|
                   \psi_{31}\rangle_{B}+|
                   \psi_{32}\rangle_{A}|
                   \psi_{32}\rangle_{B}+
                   |\psi_{33}\rangle_{A}|
                   \psi_{33}\rangle_{B}+
                   |\psi_{34}\rangle_{A}
                   |\psi_{34}\rangle_{B}\nonumber\\
                   +|\psi_{41}\rangle_{A}|
                   \psi_{41}\rangle_{B}+|
                   \psi_{42}\rangle_{A}|\psi_{42}\rangle_{B}+
                   |\psi_{43}\rangle_{A}|\psi_{43}\rangle_{B}+
                   |\psi_{44}\rangle_{A}|\psi_{44}\rangle_{B}),
\label{shi4}
\end{eqnarray}
where
\begin{eqnarray}
 |\psi_{11}\rangle_{A}=\frac{1}{2}(|\omega_{11}
 \rangle|a_{11}\rangle+|\omega_{12}\rangle
 |a_{12}\rangle+|\omega_{21}\rangle|a_{21}
 \rangle+|\omega_{22}\rangle|a_{22}\rangle);\nonumber\\
 |\psi_{12}\rangle_{A}=\frac{1}{2}(|\omega_{11}
 \rangle|a_{11}\rangle-|\omega_{12}\rangle
 |a_{12}\rangle-|\omega_{21}\rangle|a_{21}
 \rangle+|\omega_{22}\rangle|a_{22}\rangle);\nonumber
\end{eqnarray}
\begin{eqnarray}
 |\psi_{13}\rangle_{A}=\frac{1}{2}
 (|\omega_{11}\rangle|a_{11}\rangle-|
 \omega_{12}\rangle|a_{12}\rangle+|
 \omega_{21}\rangle|a_{21}\rangle-|
 \omega_{22}\rangle|a_{22}\rangle);\nonumber\\
 |\psi_{14}\rangle_{A}=\frac{1}{2}
 (|\omega_{11}\rangle|a_{11}\rangle+
 |\omega_{12}\rangle|a_{12}\rangle-
 |\omega_{21}\rangle|a_{21}\rangle-
 |\omega_{22}\rangle|a_{22}\rangle);\nonumber
\end{eqnarray}
\begin{eqnarray}
 |\psi_{21}\rangle_{A}=\frac{1}{2}
 (|\omega_{11}\rangle|a_{12}\rangle+
 |\omega_{12}\rangle|a_{21}\rangle+
 |\omega_{21}\rangle|a_{22}\rangle+
 |\omega_{22}\rangle|a_{11}\rangle);\nonumber\\
 |\psi_{22}\rangle_{A}=\frac{1}{2}
 (|\omega_{11}\rangle|a_{12}\rangle-
 |\omega_{12}\rangle|a_{21}\rangle-
 |\omega_{21}\rangle|a_{22}\rangle+
 |\omega_{22}\rangle|a_{11}\rangle);\nonumber
 \end{eqnarray}
\begin{eqnarray}
 |\psi_{23}\rangle_{A}=\frac{1}{2}
 (|\omega_{11}\rangle|a_{12}\rangle-
 |\omega_{12}\rangle|a_{21}\rangle+
 |\omega_{21}\rangle|a_{22}\rangle-
 |\omega_{22}\rangle|a_{11}\rangle);\nonumber\\
 |\psi_{24}\rangle_{A}=\frac{1}{2}
 (|\omega_{11}\rangle|a_{12}\rangle+
 |\omega_{12}\rangle|a_{21}\rangle-
 |\omega_{21}\rangle|a_{22}\rangle-|
 \omega_{22}\rangle|a_{11}\rangle);\nonumber
 \end{eqnarray}
\begin{eqnarray}
 |\psi_{31}\rangle_{A}=\frac{1}{2}
 (|\omega_{11}\rangle|a_{21}\rangle+
 |\omega_{12}\rangle|a_{22}\rangle+
 |\omega_{21}\rangle|a_{11}\rangle+
 |\omega_{22}\rangle|a_{12}\rangle);\nonumber\\
 |\psi_{32}\rangle_{A}=\frac{1}{2}
 (|\omega_{11}\rangle|a_{21}\rangle-
 |\omega_{12}\rangle|a_{22}\rangle-
 |\omega_{21}\rangle|a_{11}\rangle+
 |\omega_{22}\rangle|a_{12}\rangle);\nonumber
\end{eqnarray}
\begin{eqnarray}
 |\psi_{33}\rangle_{A}=\frac{1}{2}(|\omega_{11}
 \rangle|a_{21}\rangle-|\omega_{12}\rangle
 |a_{22}\rangle+|\omega_{21}\rangle|a_{11}
 \rangle-|\omega_{22}\rangle|a_{12}\rangle);\nonumber\\
 |\psi_{34}\rangle_{A}=\frac{1}{2}
 (|\omega_{11}\rangle|a_{21}\rangle+
 |\omega_{12}\rangle|a_{22}\rangle-|\omega_{21}\rangle
 |a_{11}\rangle-|\omega_{22}\rangle|a_{12}\rangle);\nonumber
\end{eqnarray}
\begin{eqnarray}
 |\psi_{41}\rangle_{A}=\frac{1}{2}(|\omega_{11}
 \rangle|a_{22}\rangle+|\omega_{12}\rangle
 |a_{11}\rangle+|\omega_{21}\rangle|a_{12}
 \rangle+|\omega_{22}\rangle|a_{21}\rangle);\nonumber\\
 |\psi_{42}\rangle_{A}=\frac{1}{2}(|\omega_{11}
 \rangle|a_{22}\rangle-|\omega_{12}\rangle
 |a_{11}\rangle-|\omega_{21}\rangle|a_{12}
 \rangle+|\omega_{22}\rangle|a_{21}\rangle);\nonumber
\end{eqnarray}
\begin{eqnarray}
 |\psi_{43}\rangle_{A}=\frac{1}{2}
 (|\omega_{11}\rangle|a_{22}\rangle-
 |\omega_{12}\rangle|a_{11}\rangle+
 |\omega_{21}\rangle|a_{12}\rangle-
 |\omega_{22}\rangle|a_{21}\rangle);\nonumber\\
 |\psi_{44}\rangle_{A}=\frac{1}{2}
 (|\omega_{11}\rangle|a_{22}\rangle+
 |\omega_{12}\rangle|a_{11}\rangle-
 |\omega_{21}\rangle|a_{12}\rangle-
 |\omega_{22}\rangle|a_{21}\rangle);\nonumber
\end{eqnarray}
\begin{eqnarray}
 |\psi_{11}\rangle_{B}=\frac{1}{2}
 (|\omega_{12}\rangle|a_{11}\rangle+
 |\omega_{11}\rangle|a_{12}\rangle+
 |\omega_{22}\rangle|a_{21}\rangle+
 |\omega_{21}\rangle|a_{22}\rangle);\nonumber\\
 |\psi_{12}\rangle_{B}=\frac{1}{2}
 (|\omega_{12}\rangle|a_{11}\rangle-
 |\omega_{11}\rangle|a_{12}\rangle-
 |\omega_{22}\rangle|a_{21}\rangle+
 |\omega_{21}\rangle|a_{22}\rangle);\nonumber
 \end{eqnarray}
\begin{eqnarray}
 |\psi_{13}\rangle_{B}=\frac{1}{2}
 (|\omega_{12}\rangle|a_{11}\rangle-
 |\omega_{11}\rangle|a_{12}\rangle+
 |\omega_{22}\rangle|a_{21}\rangle-
 |\omega_{21}\rangle|a_{22}\rangle);\nonumber\\
 |\psi_{14}\rangle_{B}=\frac{1}{2}
 (|\omega_{12}\rangle|a_{11}\rangle+
 |\omega_{11}\rangle|a_{12}\rangle-
 |\omega_{22}\rangle|a_{21}\rangle-
 |\omega_{21}\rangle|a_{22}\rangle);\nonumber
 \end{eqnarray}
\begin{eqnarray}
|\psi_{21}\rangle_{B}=\frac{1}{2}
(|\omega_{12}\rangle|a_{12}\rangle+
|\omega_{11}\rangle|a_{21}\rangle+
|\omega_{22}\rangle|a_{22}\rangle+
|\omega_{21}\rangle|a_{11}\rangle);\nonumber\\
|\psi_{22}\rangle_{B}=\frac{1}{2}
(|\omega_{12}\rangle|a_{12}\rangle-
|\omega_{11}\rangle|a_{21}\rangle-
|\omega_{22}\rangle|a_{22}\rangle+
|\omega_{21}\rangle|a_{11}\rangle);\nonumber
 \end{eqnarray}
\begin{eqnarray}
|\psi_{23}\rangle_{B}=\frac{1}{2}(|\omega_{12}
\rangle|a_{12}\rangle-|\omega_{11}\rangle
|a_{21}\rangle+|\omega_{22}\rangle|a_{22}
\rangle-|\omega_{21}\rangle|a_{11}\rangle);\nonumber\\
 |\psi_{24}\rangle_{B}=\frac{1}{2}
 (|\omega_{12}\rangle|a_{12}\rangle+
 |\omega_{11}\rangle|a_{21}\rangle-
 |\omega_{22}\rangle|a_{22}\rangle-
 |\omega_{21}\rangle|a_{11}\rangle);\nonumber
 \end{eqnarray}
\begin{eqnarray}
 |\psi_{31}\rangle_{B}=\frac{1}{2}(|\omega_{12}\rangle
 |a_{21}\rangle+|\omega_{11}\rangle|a_{22}\rangle+
 |\omega_{22}\rangle|a_{11}\rangle+|\omega_{21}
 \rangle|a_{12}\rangle);\nonumber\\
 |\psi_{32}\rangle_{B}=\frac{1}{2}(|\omega_{12}
 \rangle|a_{21}\rangle-|\omega_{11}\rangle|a_{22}
 \rangle-|\omega_{22}\rangle|a_{11}\rangle+|
 \omega_{21}\rangle|a_{12}\rangle);\nonumber
 \end{eqnarray}
\begin{eqnarray}
 |\psi_{33}\rangle_{B}=\frac{1}{2}(|\omega_{12}
 \rangle|a_{21}\rangle-|\omega_{11}\rangle
 |a_{22}\rangle+|\omega_{22}\rangle|a_{11}
 \rangle-|\omega_{21}\rangle|a_{12}\rangle);\nonumber\\
 |\psi_{34}\rangle_{B}=\frac{1}{2}
 (|\omega_{12}\rangle|a_{21}\rangle+|
 \omega_{11}\rangle|a_{22}\rangle-|
 \omega_{22}\rangle|a_{11}\rangle-|
 \omega_{21}\rangle|a_{12}\rangle);\nonumber
 \end{eqnarray}
\begin{eqnarray}
 |\psi_{41}\rangle_{B}=\frac{1}{2}(|
 \omega_{12}\rangle|a_{22}\rangle+|
 \omega_{11}\rangle|a_{11}\rangle+|
 \omega_{22}\rangle|a_{12}\rangle+|
 \omega_{21}\rangle|a_{21}\rangle);\nonumber\\
 |\psi_{42}\rangle_{B}=\frac{1}{2}(|
 \omega_{12}\rangle|a_{22}\rangle-|
 \omega_{11}\rangle|a_{11}\rangle-|
 \omega_{22}\rangle|a_{12}\rangle+|
 \omega_{21}\rangle|a_{21}\rangle);\nonumber
 \end{eqnarray}
\begin{eqnarray}
 |\psi_{43}\rangle_{B}=\frac{1}{2}(|
 \omega_{12}\rangle|a_{22}\rangle-|
 \omega_{11}\rangle|a_{11}\rangle+|
 \omega_{22}\rangle|a_{12}\rangle-|
 \omega_{21}\rangle|a_{21}\rangle);\nonumber\\
 |\psi_{44}\rangle_{B}=\frac{1}{2}(|
 \omega_{12}\rangle|a_{22}\rangle+|
 \omega_{11}\rangle|a_{11}\rangle-|
 \omega_{22}\rangle|a_{12}\rangle-|
 \omega_{21}\rangle|a_{21}\rangle).
\label{shi5}
\end{eqnarray}

Consequently, in this way the hyperentanglement
in frequency and
spatial mode of the two photons has been expressed as the sum of 16
tensor products of each single photon's frequency-spatial mode
entangled state. At the site of Alice, Alice can carry out $\sigma$
local operation on the spatial mode of photon A. There are 16 local
operations, and each corresponds to an encoding, i.e.
\begin{eqnarray}
\sigma_{1}\leftrightarrow(0000); \sigma_{2}
\leftrightarrow(0001);\sigma_{3}
\leftrightarrow(0010);\sigma_{4}
\leftrightarrow(0011);\nonumber\\
\sigma_{5}\leftrightarrow(0100);
\sigma_{6}\leftrightarrow(0101);
\sigma_{7}\leftrightarrow(0110);
\sigma_{8}\leftrightarrow(0111);\nonumber\\
\sigma_{9}\leftrightarrow(1000);
\sigma_{10}\leftrightarrow(1001);
\sigma_{11}\leftrightarrow(1010);
\sigma_{12}\leftrightarrow(1011);\nonumber\\
\sigma_{13}\leftrightarrow(1100);
\sigma_{14}\leftrightarrow(1101);
\sigma_{15}\leftrightarrow(1110);
\sigma_{16}\leftrightarrow(1111).
\label{shi6}
\end{eqnarray}
The results of these 16 local operations implemented on spatial mode
$|\psi\rangle_{S}=(|a_{11}\rangle|b_{11}\rangle
+|a_{12}\rangle|b_{12}\rangle+|a_{21}
\rangle|b_{21}\rangle+|a_{22}\rangle|b_{22}\rangle)$ are
\begin{eqnarray}
\sigma_{1}£º|\psi\rangle_{S}\rightarrow|
\psi_{1}\rangle_{S}=(|a_{11}\rangle|b_{11}\rangle
+|a_{12}\rangle|b_{12}\rangle+|a_{21}\rangle|b_{21}
\rangle+|a_{22}\rangle|b_{22}\rangle);\nonumber\\
\sigma_{2}£º|\psi\rangle_{S}\rightarrow|\psi_{2}
\rangle_{S}=(|a_{11}\rangle|b_{11}\rangle-|a_{12}
\rangle|b_{12}\rangle-|a_{21}\rangle|b_{21}
\rangle+|a_{22}\rangle|b_{22}\rangle);\nonumber
 \end{eqnarray}
\begin{eqnarray}
\sigma_{3}£º|\psi\rangle_{S}\rightarrow|\psi_{3}
\rangle_{S}=(|a_{11}\rangle|b_{11}\rangle-|a_{12}
\rangle|b_{12}\rangle+|a_{21}\rangle|b_{21}
\rangle-|a_{22}\rangle|b_{22}\rangle);\nonumber\\
\sigma_{4}£º|\psi\rangle_{S}\rightarrow|\psi_{4}
\rangle_{S}=(|a_{11}\rangle|b_{11}\rangle+|a_{12}
\rangle|b_{12}\rangle-|a_{21}\rangle|b_{21}
\rangle-|a_{22}\rangle|b_{22}\rangle);\nonumber
 \end{eqnarray}
\begin{eqnarray}
\sigma_{5}£º|\psi\rangle_{S}\rightarrow|
\psi_{5}\rangle_{S}=(|a_{11}\rangle|b_{12}
\rangle+|a_{12}\rangle|b_{21}\rangle+|a_{21}
\rangle|b_{22}\rangle+|a_{22}\rangle|b_{11}
\rangle);\nonumber\\
\sigma_{6}£º|\psi\rangle_{S}\rightarrow|
\psi_{6}\rangle_{S}=(|a_{11}\rangle|b_{12}
\rangle-|a_{12}\rangle|b_{21}\rangle-|a_{21}
\rangle|b_{22}\rangle+|a_{22}\rangle|b_{11}\rangle);\nonumber
 \end{eqnarray}
\begin{eqnarray}
\sigma_{7}£º|\psi\rangle_{S}\rightarrow|\psi_{7}
\rangle_{S}=(|a_{11}\rangle|b_{12}\rangle-|a_{12}
\rangle|b_{21}\rangle+|a_{21}\rangle|b_{22}
\rangle-|a_{22}\rangle|b_{11}\rangle);\nonumber\\
\sigma_{8}£º|\psi\rangle_{S}\rightarrow|
\psi_{8}\rangle_{S}=(|a_{11}\rangle|b_{12}
\rangle+|a_{12}\rangle|b_{21}\rangle-|a_{21}
\rangle|b_{22}\rangle-|a_{22}\rangle|b_{11}\rangle);\nonumber
 \end{eqnarray}
\begin{eqnarray}
\sigma_{9}£º|\psi\rangle_{S}\rightarrow|\psi_{9}
\rangle_{S}=(|a_{11}\rangle|b_{21}\rangle+|a_{12}
\rangle|b_{22}\rangle+|a_{21}\rangle|b_{11}
\rangle+|a_{22}\rangle|b_{12}\rangle);\nonumber\\
\sigma_{10}£º|\psi\rangle_{S}\rightarrow|
\psi_{10}\rangle_{S}=(|a_{11}\rangle|b_{21}
\rangle-|a_{12}\rangle|b_{22}\rangle-|a_{21}
\rangle|b_{11}\rangle+|a_{22}\rangle|b_{12}\rangle);\nonumber
 \end{eqnarray}
\begin{eqnarray}
\sigma_{11}£º|\psi\rangle_{S}\rightarrow|
\psi_{11}\rangle_{S}=(|a_{11}\rangle|b_{21}
\rangle-|a_{12}\rangle|b_{22}\rangle+|a_{21}
\rangle|b_{11}\rangle-|a_{22}\rangle|b_{12}
\rangle);\nonumber\\
\sigma_{12}£º|\psi\rangle_{S}\rightarrow|
\psi_{12}\rangle_{S}=(|a_{11}\rangle|b_{21}
\rangle+|a_{12}\rangle|b_{22}\rangle-|a_{21}
\rangle|b_{11}\rangle-|a_{22}\rangle|b_{12}\rangle);\nonumber
 \end{eqnarray}
\begin{eqnarray}
\sigma_{13}£º|\psi\rangle_{S}\rightarrow|\psi_{13}
\rangle_{S}=(|a_{11}\rangle|b_{22}\rangle+|a_{12}
\rangle|b_{11}\rangle+|a_{21}\rangle|b_{12}\rangle+
|a_{22}\rangle|b_{21}\rangle);\nonumber\\
\sigma_{14}£º|\psi\rangle_{S}\rightarrow|\psi_{14}
\rangle_{S}=(|a_{11}\rangle|b_{22}\rangle-|a_{12}
\rangle|b_{11}\rangle-|a_{21}\rangle|b_{12}\rangle+
|a_{22}\rangle|b_{21}\rangle);\nonumber
 \end{eqnarray}
\begin{eqnarray}
\sigma_{15}£º|\psi\rangle_{S}\rightarrow|\psi_{15}
\rangle_{S}=(|a_{11}\rangle|b_{22}\rangle-|a_{12}
\rangle|b_{11}\rangle+|a_{21}\rangle|b_{12}\rangle-
|a_{22}\rangle|b_{21}\rangle);\nonumber\\
\sigma_{16}£º|\psi\rangle_{S}\rightarrow|\psi_{16}
\rangle_{S}=(|a_{11}\rangle|b_{22}\rangle+|a_{12}
\rangle|b_{11}\rangle-|a_{21}\rangle|b_{12}
\rangle-|a_{22}\rangle|b_{21}\rangle).
\label{shi7}
\end{eqnarray}
Thus corresponding to the 16 local operations, the initial state
$|\psi\rangle$ will be turned into 16 kinds of form, for instance,
the state (3) corresponds to operation $\sigma_{1}$.

After local operation, Alice measures the frequency-spatial mode
entangled state of her photon by virtue of the measurement setup
shown in appendix. Take the state (4) for example, in case the
result obtained by Alice is $|\psi_{11}\rangle_{A}$, she tells Bob
her result via classical communication. Depending on Alice's
measurement result, Bob's photon will be projected to the
corresponding state $|\psi_{11}\rangle_{B}$. Bob subsequently
measures his photon with the same setup, his result will be
$|\psi_{11}\rangle_{B}$ which is perfectly correlated to Alice's
result. Thereupon, according to Alice's result and his result, Bob
can be well aware that the state after local operation must be the
state (4), and the corresponding local operation is $\sigma_{1}$. In
this way, Alice and Bob can take the encoding of $\sigma_{1}$ as
their determinate secure key, and the $|\psi_{11}\rangle_{B}$ as
random secure key.

In theory, this protocol can absolutely generate a four-bit determinate
secure key and a four-bit random secure key for one pair of photon. And
so the QKD protocol based on high dimensional hyperentangled state
gives rise to a larger coding density and greater security against
eavesdropping attacks.

\section{Conclusions}

In conclusion, we have demonstrated a novel scheme for the
preparation of a two-photon high dimensional hyperentangled state.
Our scheme can encode the state of the two photons into
two-dimensional, four-dimensional and four-dimensional entanglement in
polarization, frequency and spatial mode respectively by dint of 
linear optical instruments and cross-Kerr nonlinearity.
Theoretically, this scheme is simply extended to higher dimensional
qudits and more photons. Furthermore, we have presented a quantum
key distribution protocol utilizing the high dimensional
hyperentangled state. This protocol is based on the
hyperentanglement swapping between frequency and spatial mode.
Compared to the normal high dimensional entangled state, the
protocol has a higher utilization rate of photon pairs. Meanwhile,
the message in our protocol is encoded via a multilevel system
which enable us obtain a larger flux of information in comparison with
the usual two-dimensional QKD protocols.

\vskip 0.7cm
\noindent {\bf Acknowledgments}

This work was supported in part by the National Natural Science
Foundation of China(NSFC) under the grants 10935012, 10821063 and
11175249.


\pagebreak

\noindent
\appendix{\noindent\bf{Appendix: Measurement
method for 4-dimensional frequency-spatial mode entangled state}}
\label{app:def}

\begin{figure}[htp]
\centering
\includegraphics[width=0.4\textwidth]{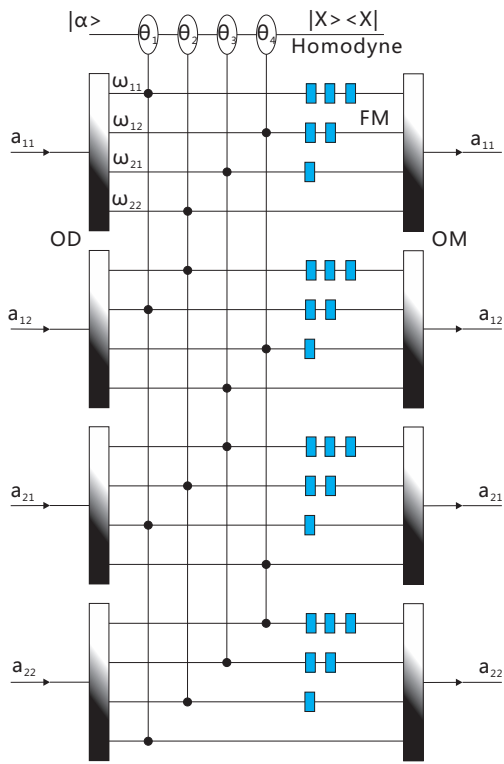}
\caption{\small The first part of measurement setup for
4-dimensional
frequency-spatial mode entangled state.} \label{fig4} \vspace{-0mm}
\end{figure}

\begin{figure}[htp]
\centering
\includegraphics[width=0.7\textwidth]{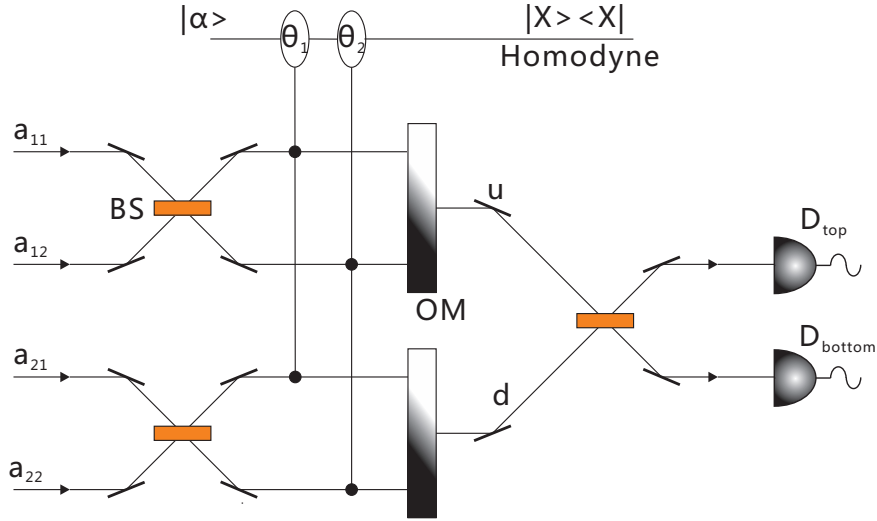}
\caption{\small The second part of measurement setup
for
4-dimensional frequency-spatial mode entangled state.}
\label{fig5} \vspace{-0mm}
\end{figure}

Taking photon A for example, we can obtain a one-to-one relationship
between the 16 states and the results of measurement by the setup
shown in Figure \ref{fig4} and \ref{fig5}. For the convenience of
paper, the 16 states are divided into the following four groups:
\begin{eqnarray}
 G_{1}:\{|\psi_{11}\rangle_{A},|\psi_{12}
 \rangle_{A},|\psi_{13}\rangle_{A},|
 \psi_{14}\rangle_{A}\},
 G_{2}:\{|\psi_{21}\rangle_{A},|\psi_{22}\rangle_{A},|
 \psi_{23}\rangle_{A},|\psi_{24}\rangle_{A}\}, \nonumber\\
 G_{3}:\{|\psi_{31}\rangle_{A},|\psi_{32}\rangle_{A},|
 \psi_{33}\rangle_{A},|\psi_{34}\rangle_{A}\},
 G_{4}:\{|\psi_{41}\rangle_{A},|\psi_{42}\rangle_{A},|
 \psi_{43}\rangle_{A},|\psi_{44}\rangle_{A}\}.
\label{shi1}
\end{eqnarray}
We can make use of bit-flip operations to implement
the conversions
between the states in the same order of different groups, and
phase-flip operations to implement the conversions between the
states of the same group.

In the first part (shown in Figure \ref{fig4}), split
every path into
four new different paths according to photon's frequency with four
optical demultiplexers (OD). Then lead the 16 paths to a cross-kerr
nonlinear medium, after measuring the phase shift of the coherent state,
we would know which incident state belongs to which group. Concretely,
the phase shifts $\theta_{1},\theta_{2},\theta_{3}$ and $\theta_{4}$
correspond to $G_{1},G_{2},G_{3},G_{4}$ respectively. Afterwards use
single-photon frequency multipliers to make the frequencies of all
paths the same, that is the information of frequency is erased.
Finally, combine the 16 paths into four paths again with four
optical multiplexers(OM) \cite{WDM}. After the first part, the
states of spatial mode corresponding to each group are in the
following form:
\begin{eqnarray}
\varphi_{1}=\frac{1}{2}(|a_{11}\rangle+|a_{12}\rangle+
|a_{21}\rangle+|a_{22}\rangle),\varphi_{2}=\frac{1}{2}
(|a_{11}\rangle-|a_{12}\rangle-|a_{21}\rangle+|a_{22}
\rangle),\nonumber\\
\varphi_{3}=\frac{1}{2}(|a_{11}\rangle-|a_{12}\rangle+
|a_{21}\rangle-|a_{22}\rangle),\varphi_{4}=\frac{1}{2}
(|a_{11}\rangle+|a_{12}\rangle-|a_{21}\rangle-|a_{22}\rangle).
\label{shi1}
\end{eqnarray}

In the second part (shown in Figure \ref{fig5}), all the
three BSs
implement the following transform:
\begin{eqnarray}
\frac{1}{\sqrt{2}}(|T\rangle+|B\rangle)\rightarrow|T\rangle;\frac{1}
{\sqrt{2}}(|T\rangle-|B\rangle)\rightarrow|B\rangle.
\label{shi1}
\end{eqnarray}
In the same manner, after two BSs, let the paths pass through
a
cross-kerr nonlinear medium. If the phase shift of coherent state is
$\theta_{1}$, that means the input state is $\varphi_{1}$ or
$\varphi_{4}$, and if the phase shift is $\theta_{2}$, the input
state should be $\varphi_{2}$ or $\varphi_{3}$. Then combine the
upper (under) two paths into path u (d), and let the paths u and d
transit a BS again. As a result, the response of upper (lower)
detector indicates the input state is $\varphi_{1}$ or $\varphi_{3}$
($\varphi_{2}$ or $\varphi_{4}$).

To sum up, Alice (Bob) can distinguish all the 16
frequency-spatial
mode entangled states via our measurement setup.

\end{document}